\documentclass[%
 reprint,
superscriptaddress,
 amsmath,amssymb,
 aps,
prl,
]{revtex4-2}

\usepackage{graphicx}
\usepackage{dcolumn}
\usepackage{bm}
\usepackage[colorlinks=true, citecolor=blue, linkcolor=blue, urlcolor=blue]{hyperref}

\usepackage{color, soul}

\begin{document}
\preprint{APS/123-QED}

\title{
Intrinsic magnetization of the superconducting condensate in Fe(Te,Se)
}

\author{Mohammad Javadi Balakan}
\email{mjavadibalakan@albany.edu}
\affiliation{College of Nanotechnology, Science, and Engineering, State University of New York, Albany, NY 12203, USA.}

\author{Shiva Heidari}
\affiliation{Physics Department, City College of the City University of New York, NY 10031, USA.}

\author{Genda Gu}
\affiliation{Condensed Matter Physics and Materials Science Division, Brookhaven National Laboratory, Upton, NY 11973, USA.}

\author{Qiang Li}
\affiliation{Condensed Matter Physics and Materials Science Division, Brookhaven National Laboratory, Upton, NY 11973, USA.}
\affiliation{Department of Physics and Astronomy, Stony Brook University; Stony Brook, New York 11794, USA.}

\author{Kenji Watanabe}
\affiliation{National Institute for Materials Science, 1-1 Namiki, Tsukuba 305-0044, Japan.}

\author{Takashi Taniguchi}
\affiliation{National Institute for Materials Science, 1-1 Namiki, Tsukuba 305-0044, Japan.}

\author{Ji Ung Lee}
\email{jlee1@albany.edu}
\affiliation{College of Nanotechnology, Science, and Engineering, State University of New York, Albany, NY 12203, USA.}

\date{\today}
\begin{abstract}
A spin-polarized superconducting condensate generates a net magnetization with measurable signatures. We present evidence for an intrinsic magnetic field in mesoscopic Fe(Te,Se) rings. The intrinsic field, encoded in the \textit{phase} of superconducting quantum oscillations, scales linearly with the DC bias current, and its orientation exhibits an anomalous dependence on polarity and magnitude of the applied current. The magnetoresistance displays a \textit{dual} flux-quantization effect with respect to the external magnetic field and the DC current. A minimal model incorporating Rashba coupling with an effective anisotropic out-of-plane interaction accounts for the experimental observations. These results provide evidence for spin-polarized superconductivity at the device scale and open new opportunities for superconducting spintronic and quantum information platforms.
\end{abstract}

\maketitle

\begin{figure}[t]
\includegraphics[width=0.48\textwidth]{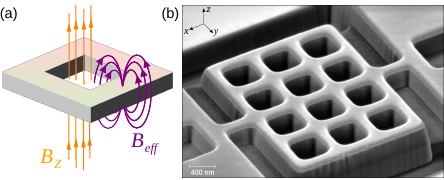}
\caption{\label{fig_sem} (a) Schematic illustration of the external field $B_z$ and the intrinsic effective field $B_\text{eff}$ contributing to the magnetic flux through a superconducting ring. (b) SEM image of a representative device. The layer stack from top to bottom is hBN/Fe(Te,Se)/SiO$_2$(300~nm)/Si.}
\end{figure}

Spin polarization in superconductivity is one of the most sought-after phenomena in condensed-matter physics, as it represents a fundamental departure from conventional spin-singlet superconductors, in which Cooper pairs form exclusively between electrons of opposite spin. Certain triplet superconductors with half-integer winding (HIW) of the order parameter are predicted to support half-quantum vortices hosting non-Abelian Majorana zero modes, a key ingredient in fault-tolerant topological quantum computing~\cite{volovik_1976,volovik_1999_fermion,readgreen_2000,ivanov_2001,kitaev_2002,nayak_2008, alicea_2012, yazdani_2023}. A direct manifestation of this nontrivial phase winding is half-integer flux quantization in annular geometries. Recently, we reported the emergence of HIW states in mesoscopic Fe(Te,Se) rings, evidenced by magnetoresistance (MR) oscillations with a periodicity of $\Phi_0/2$, where $\Phi_0=h/2e$ is the flux quantum~\cite{balakan_hiw}. The appearance of these exotic states implies that the order parameter contains a triplet pairing component and, importantly, carries a finite spin polarization, which may lead to an intrinsic magnetization of the condensate. The spin polarization associated with a stable HIW state is given by $S=(g_s\mu_B)^{-1}\chi \mathcal{B}$, where the total magnetic field $\mathcal{B}\!=\!B_z\!+\!B_\text{eff}$ consists of an external field ($B_z$) and an intrinsic effective field ($B_\text{eff}$) arising from the spin currents~\cite{vakaryuk_2009}. This effective field develops independently of the external field and can contribute to the net magnetic flux threading a superconducting ring [Fig.~\ref{fig_sem}(a)]. Accordingly, for a spin-polarized superconducting condensate, flux quantization generalizes to $n\Phi_0\!=\!(B_z\!+\!B_\text{eff})\mathcal{A}$, rendering the intrinsic field detectable in Little–Parks (LP) experiment~\cite{littlepark_1962,littlepark_1964}, as we show in this letter.

Here, we present evidence for intrinsic magnetization of the superconducting condensate in mesoscopic Fe(Te,Se) rings. The associated effective field manifests through a bias-dependent phase modulation of the quantum oscillations. We observe a dual flux-quantization effect. In addition to the well-known LP oscillations under an external magnetic field, the oscillations also occur as a function of the applied DC current, confirming the intrinsic nature of the effective field. Our studies show that while the effective field scales inversely with the ring size, the electronic magnetization coefficient remains uniform across different samples. These results provide strong evidence for spin-polarized superconductivity in mesoscopic Fe(Te,Se) devices, offering additional support for the emergence of HIW states in this material system.

\begin{figure*}[t]
\includegraphics[width=0.97\textwidth]{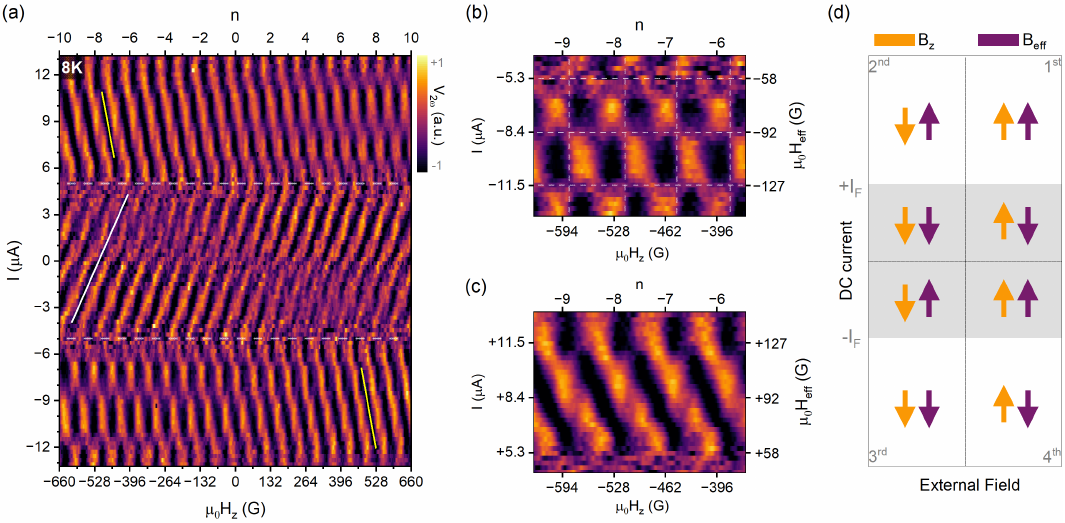}
\caption{\label{fig_heff} (a) Normalized MR as a function of external field and DC bias current at 8~K. The winding number is defined as $n = B_z / \Delta B_z$, with $\Delta B_z\approx66$~G. The white and yellow lines are constant-phase trajectories that illustrate the gradual shift of the LP oscillations with bias current. Dashed horizontal lines mark $I_F$, the bias current at which the polarity of $B_\text{eff}$ reverses. (b, c) Enlarged MR maps from the third and second quadrants of panel a, respectively. The effective field is determined from the slope of constant-phase lines. (d) Schematic illustrating the relative orientation of the external ($B_z$) and intrinsic ($B_\text{eff}$) fields in the four quadrants of panel a.} 
\end{figure*}

The device configuration follows our previous report and consists of an array of identical Fe(Te$_{0.55}$,Se$_{0.45}$) rings fully encapsulated by a thin hBN layer [Fig.~\ref{fig_sem}(b)]. Transport measurements are performed using a four-terminal AC+DC method, where a $1~\mu \text{A}$ AC excitation is superimposed on a variable DC current to probe the phase space of the nonlinear (second-harmonic) MR. This approach provides direct insight into the winding nature of the superconducting order parameter. Details of the multi-ring structure, device fabrication, measurement circuitry, and the nonlinear MR method are described in our previous report~\cite{balakan_hiw}. In the following, we present results for device 1, which shows a superconducting transition at 13.3~K and becomes non-resistive below 7.5~K. The effective area of each ring is $\mathcal{A}\!=\!0.29\pm0.03~\mu\text{m}^2$, corresponding to a flux quantum of $\Delta B\!=\!72\pm7$~G. Similar transport characteristics are observed across multiple devices with various ring sizes.

\begin{figure*}[t]
\includegraphics[width=0.98\textwidth]{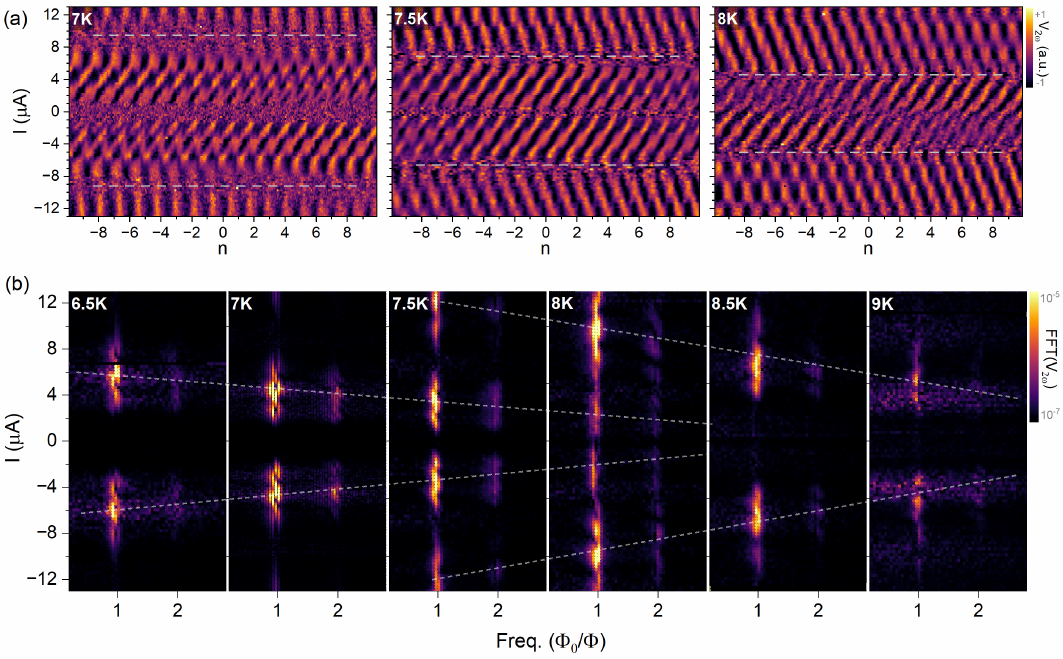}
\caption{\label{fig_temperature}  (a) MR spectra over a range of temperatures near $T_c$. Dashed lines mark the characteristic currents $I_F$. (b) FFT of the MR spectra over a wide temperature range. The FFT peaks at $\Phi_0/\Phi=1~(2)$ correspond to the IW (HIW) of the order parameter. The inner (outer) dashed lines trace the quantum oscillations associated with $\alpha_-$ ($\alpha_+$) coupling regimes.}
\end{figure*}

Figure~\ref{fig_heff}(a) presents the nonlinear MR as a function of external magnetic field and DC bias current at 8~K. The spectrum displays a hierarchy of symmetries together with a complex bias-dependent structure. Among the prominent features are MR oscillations with a periodicity of $\Delta B_z\!\approx\!66$~G, consistent with the flux quantum set by the ring geometry. These oscillations originate from the integer winding (IW) of the order parameter and therefore reflect Cooper-pair flux quantization. At low DC current and low magnetic field, the IW modes undergo a distinct splitting that gives rise to HIW states, signaling a nontrivial phase winding characteristic of topological superconductors. Beyond these winding signatures, the MR oscillations also exhibit a global current–field symmetry. The spectrum remains invariant \textit{only} under concurrent inversion of the magnetic field and DC bias current, $V_{2\omega}(I,B_z)\!=\!- V_{2\omega}(-I,-B_z)$, indicating the presence of strong spin-orbit coupling (SOC) in the superconducting condensate. A detailed discussion of these features is provided in Ref.~\cite{balakan_hiw}.

A key observation of this letter is that the \textit{phase of quantum oscillations} evolves with the DC bias. The MR oscillations shown in Fig.~\ref{fig_heff}(a) exhibit a pattern of gradual shift with the applied current, implying a bias-dependent phase modulation as 
\begin{equation}
    \label{eq:qphase}
    \Delta V=V_0\ cos\!\left[\dfrac{2\pi}{\Phi_0}\left(\Phi+\theta_\text{LP}\right)\right]~\!,
\end{equation}
where $V_0$ is the amplitude of oscillations and $\Phi=B_z\mathcal{A}$ is the applied flux (see \ \textcolor{cyan}{Appendix A}). Here, $\theta_\text{LP}$ is the phase of Little--Parks oscillations, which is independent of the external field and varies linearly with the applied current. This behavior points to the presence of an intrinsic magnetic field with an effective out-of-plane component $B_\text{eff}$ expressed as
\begin{equation}
    \label{eq:b_eff}
    \theta_\text{LP} = B_\text{eff}\mathcal{A} = \alpha I \mathcal{A}~\!,
\end{equation}
where $\alpha$ is the magnetoelectric coupling strength, which can be determined from the slope of constant-phase trajectories $\alpha\!=\!-d B_z/dI$ as indicated in Fig.~\ref{fig_heff}(a). Although $B_\text{eff}$ varies linearly with the bias current, it is distinct from the classical Oersted field. In a ring geometry, the current flows through both arms, causing the magnetic fields from the two branches to nearly cancel inside the ring. Even assuming a fully circulating DC current, numerical integration shows that more than 1~mA would be required to produce a single flux quantum, which is orders of magnitude larger than the currents used here. Furthermore, the orientation of $B_\text{eff}$ depends anomalously on \textit{both polarity and magnitude} of the DC current. As indicated in Fig.~\ref{fig_heff}(a), two distinct regimes emerge: at $|I|\!<\!I_F\! \approx\! 5~\mu\text{A}$, the coupling factor is negative $\alpha_{_-} \!\approx \!-26~\text{G}/\mu\text{A}$ (white line), whereas for $|I|\!>\!I_F$ it becomes positive $\alpha_{_+}\! \approx\!+11~\text{G}/\mu\text{A}$ (yellow lines). This implies that $B_\text{eff}$ reverses abruptly at $I_F\approx\pm5~\mu\text{A}$. For instance, for DC currents between 0 and $5~\mu\text{A}$, the effective field points along $-\hat{z}$, while for $I\! >\! 5~\mu\text{A}$, $B_\text{eff}$ flips to $+\hat{z}$. This sign change occurs without reversing the current direction, demonstrating that $B_\text{eff}$ is fundamentally inconsistent with Ampère’s law.

The phase modulation of superconducting flux quantum oscillations presents a new effect that, to our knowledge, has not been previously observed in superconducting rings. In fluxoid quantization experiments on both conventional and unconventional superconductors, including single-ring~\cite{cai_2013,li_2019,yasui_2020,xu_2020,cai_2022,wan_2024, xu_2024, Ge_2024} and multi-ring~\cite{tinkham_1983,ooi_2005,sochnikov_2010,avci_2010,berdiyorov_2012,bi_2025} devices, the LP phase is fixed (0 or $\pi$) and remains unchanged as the bias current or temperature is varied. Our control experiments on identical 12-ring devices based on a conventional $s$-wave superconductor (NbTiN) display featureless LP oscillatory spectrum without any detectable phase modulation.

Interestingly, the emergence of an effective intrinsic field, in addition to the externally applied magnetic field, gives rise to a \textit{dual} flux-quantization effect, manifested as a checkerboard pattern in the first ($I>0,, H_z>0$) and third ($I<0,, H_z<0$) quadrants of Fig.~\ref{fig_heff}(a). In the presence of $B_\text{eff}$, the total magnetic field experienced by a ring becomes $\mathcal{B}\!=\!B_z\!+\!B_\text{eff}$, and the quantization condition generalizes to $n\Phi_0 = (B_z + B_\text{eff})\mathcal{A}$ [Eq.~\ref{eq:qphase}]. As shown in Fig.~\ref{fig_heff}(b), sweeping $B_z$ at a fixed $B_\text{eff}$ (constant $I$) produces horizontal MR oscillations with a periodicity of $\Delta B_z \approx 66$~G, corresponding to one flux quantum. Conversely, sweeping the DC current at a fixed external field generates vertical oscillations with a half-period of $\Delta I\!\approx\!3.1\pm0.3~\mu\text{A}$, which corresponds to $\Delta B_\text{eff}\!=\!\alpha_{_+}(2\Delta I)\!\approx\!68$~G, again matching a single quantum of flux. Such a dual-quantization further confirms that $B_\text{eff}$ is independent of the external field.

The complex structure of the MR spectrum is dictated by the relative orientation of $B_z$ and $B_\text{eff}$. When the two fields are parallel, the MR spectrum develops a checkerboard pattern shown in Fig.~\ref{fig_heff}(b). In contrast, an antiparallel alignment produces a drift pattern highlighted in Fig.~\ref{fig_heff}(c). For $|I| \!>\! I_F$, the parallel alignment occupies the first and third quadrants of Fig.~\ref{fig_heff}(a), while the antiparallel alignment appears in the second and fourth quadrants. Below the flip current, $|I| \!< \!I_F$, this configuration reverses in such a way that the parallel (antiparallel) alignment appears in the second and fourth (first and third) quadrants. A schematic of these orientations is shown in Fig.~\ref{fig_heff}(d). 

\begin{figure}[t]
\includegraphics[width=0.48\textwidth]{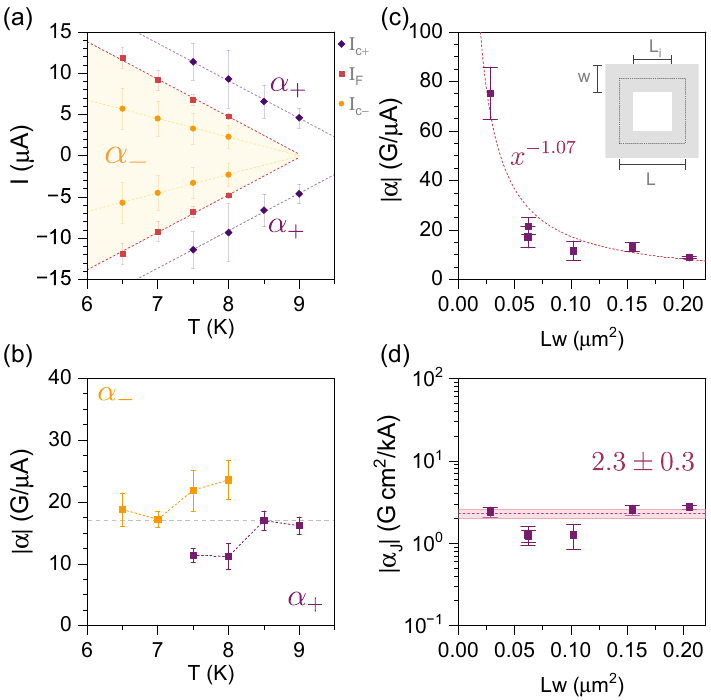}
\caption{ \label{fig_coupling} (a) Phase diagram of coupling factor. Dashed lines are linear fits. Magnetoelectric coupling strength $|\alpha_\pm|$ as a function of (b) temperature and (c) ring size. The latter is collected over multiple devices. Inset in c shows the ring geometry. (d) Magnetoelectric coefficient per current-density as function of ring size.}
\end{figure}

The abrupt reversal of $B_\text{eff}$ at $I_F$ points to an unconventional coupling mechanism between the effective field and the DC current. Figure~\ref{fig_temperature}(a) shows MR spectra over a range of temperatures near $T_c$. The two sets of LP oscillations with opposite couplings are consistently observed across a range of temperatures, reflecting the same intrinsic field behavior described earlier. While $I_F$ decreases with increasing temperature, the coupling factor remains negative (positive) for $|I|\!<\!I_F$ ($|I|\!>\!I_F$). Therefor, the relative alignment of $B_\text{eff}$ and $B_z$ shown in Fig.~\ref{fig_heff}(d) maintains the same configuration regardless of temperature. 

The emergence of two opposite coupling factors $\alpha_\pm$ is consistent with a scenario of two Fermi surfaces FS$^\pm$ with slightly different transition temperatures (critical currents). Their distinct critical currents are evident in Fig.~\ref{fig_temperature}(b), where the Fast Fourier Transform (FFT) of the MR spectra is presented over a broader temperature range. For example, at 7.5 K, two strong FFT signals appear at $\pm4~\mu\text{A}$ and $\pm12~\mu\text{A}$, which can be assigned to the critical currents of FS$^-$ ($I_c^{-}$) and FS$^+$ ($I_c^{+}$), respectively. Both characteristic currents decrease with temperature, as traced by the white dashed lines. Within the bias window shown in Fig.~\ref{fig_temperature}(b), the FFT spectrum is mostly governed by FS$^-$ at $T\!\le\!7$~K, and by FS$^+$ at $T\!\ge\!8.5$~K, while in the intermediate temperature range both Fermi surfaces contribute to the spectrum.

\begin{figure}[t]
\includegraphics[width=0.48\textwidth]{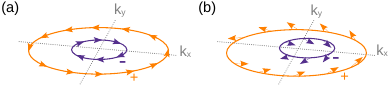}
\caption{ \label{fig_SOC} Schematics of two concentric Fermi surfaces with opposite spin–momentum locking. (a) Rashba SOC in equilibrium, where the opposite helicity bands carry in-plane spin textures. (b) Rashba SOC with an effective out-of-plane interaction $g_z(\mathbf{k})$ in the presence of a DC current. The current drives a non-equilibrium imbalance and $g_z(\mathbf{k})$ cants the spin textures to generate opposite $s_z(\mathbf{k})$ components.}
\end{figure}

In the above picture, $I_F$ marks the DC current below (above) which the quantum oscillations are governed by FS$^-$ (FS$^+$). Figure~\ref{fig_coupling}(a) shows the resulting phase diagram determined from the temperature dependence of the characteristic currents $I_c^{-}$, $I_F$, and $I_c^{+}$. Over the measured range, $I_F$ and $I_c^{\pm}$ all vary linearly with temperature. While $I_c^{-}$ decreases at a rate of $-2.25~\mu\text{A/K}$, both $I_F$ and $I_c^{+}$ exhibit the same slope of $-4.5~\mu\text{A/K}$, indicating that the flip current is essentially set by the FS$^+$. Temperature dependence of the magnetoelectric coupling magnitude is presented Fig.~\ref{fig_coupling}(b). While both $\alpha_{-}$ and $\alpha_{+}$ increase modestly with temperature, the average magnitude of $|\alpha|$ remains approximately constant across the measured range.

We have observed similar intrinsic magnetization across multiple Fe(Te,Se) devices with different ring sizes (see \textcolor{cyan}{Supplemental Material (SM)}~\cite{sm}). Figure~\ref{fig_coupling}(c) consolidates our results across different samples where the coupling factor is plotted as a function of the ring size Lw (the area over which the superfluid circulates). The data reveal a robust inverse scaling with respect to the ring size following $\alpha\sim(\text{Lw})^{-1.07\pm0.13}$ (see \textcolor{cyan}{Appendix B}). The electronic magnetization, however, is expected to reflect a fundamental property inherent to the material system. Expressing the intrinsic field in terms of current density, $B_\text{eff}=\alpha_{_J}J$, isolates the magnetoelectric coefficient $\alpha_J$ that should be independent of the device geometry. As demonstrated in Fig.~\ref{fig_coupling}(d), $\alpha_J$ remains essentially constant across multiple devices leading to an average value of $\alpha_J\!=\!2.3\pm0.3~\text{G}/\text{(kA/cm}^{2})$. Notably, this value is comparable to the current-induced electronic magnetization of $2.6\pm0.6~\text{G}/(\text{kA}/\text{cm}^2)$ associated with one Te atom per unit cell in bulk tellurium, where the magnetization is also found to scale linearly with the applied current~\cite{furukawa_2017}.\\

\paragraph*{Theory:} The global current–field symmetry of the MR oscillations together with the observation of two opposite magnetoelectric coefficients $\alpha_\pm$, suggest the presence of two Fermi surfaces with opposite spin textures and slightly different transition temperatures. A natural description is Rashba-type spin–orbit coupling, which produces two concentric, helicity-resolved Fermi surfaces [Fig.~\ref{fig_SOC}(a)]. Such coupling may arise from the quasi-two-dimensional nature of superconductivity in Fe(Te,Se) mesoscopic devices, as reported previously~\cite{lin_2015,tang_2019,lin_2023} and supported by our own experiments (data not shown). Angle-resolved photoemission studies have also revealed signatures of spin–momentum locking in this material system~\cite{zhang_2018,zhang_2019,zhang_2021}. In Rashba-split bands, the gap function acquires mixed spin-singlet and spin-triplet components, and a DC current can induce finite spin polarization via the Edelstein effect~\cite{Edelstein_1989,Edelstein_1995}. This polarization, however, is confined to the transport plane~\cite{fujimoto_2007}. Our results can be accounted for by incorporating an effective anisotropic out-of-plane spin–orbit interaction. In such a system, Fermi surfaces of opposite helicity ($\mathrm{FS}^{\pm}$) acquire opposite spin-z components. A DC current generates a non-equilibrium imbalance, resulting in a net out-of-plane magnetization [Fig.~\ref{fig_SOC}(b)]. This mechanism accounts for the observed current-driven intrinsic magnetization, which remains independent of the external field (see \textcolor{cyan}{Appendix C}).

The details of the theoretical analysis are presented in the \textcolor{cyan}{Supplemental Material}. Here we outline the essential elements of a minimal model that captures the fundamental relation governing the coupling factor. The Hamiltonian of a 2D system with both in-plane and out-of-plane SOC can be written as
\begin{equation}
H(\mathbf{k})=\xi_{\mathbf{k}}\sigma_0+\mathbf{g}(\mathbf{k})\cdot\boldsymbol{\sigma},
\label{eq:H_gsigma}
\end{equation}
where $\mathbf{g}(\mathbf{k})\!=\!\left(\lambda_R (-k_y,k_x),\;g_z(\mathbf{k})\right)$ is the spin-orbit field. Here $\boldsymbol{\sigma}$ denotes the vector of Pauli matrices, $\lambda_R$ is the Rashba coupling strength, and $g_z(\mathbf{k})$ represents the out-of-plane component. The dispersion and spin texture of the bands are given by $E_{\mathbf{k},s}\!=\!\xi_{\mathbf{k}}\!+\!s|\mathbf{g}(\mathbf{k})|$ and $\langle\boldsymbol{\sigma}\rangle_s(\mathbf{k})\!=\!s\,\hat{\mathbf{g}}(\mathbf{k})$, respectively, where $s\!=\!\pm$ labels the helicity. The spin texture on the two helicity-resolved Fermi surfaces acquires opposite out-of-plane canting, accounting for the observed sign reversal of $B_\text{eff}$ shown in Fig.~\ref{fig_heff}(d). A uniform supercurrent corresponds to a condensate phase gradient of $\mathbf{Q}$ (pair momentum), which Doppler-shifts quasiparticle energies by
$\delta E_{\mathbf{k},s}\simeq \hbar\,\mathbf{v}_{\mathbf{k},s}\cdot\mathbf{Q}$ and results in occupation change of
$\delta n_{\mathbf{k},s}\simeq-(\partial f/\partial E)\,\delta E_{\mathbf{k},s}$ near the Fermi surface. The resulting out-of-plane magnetization (e.g. for FS$^+$) is given by $M_z(\mathbf{Q})\!=\!\kappa\,\mathbf{Q}$, where $\kappa$ is the magnetoelectric kernel, a Fermi-surface functional determined by $\hat{\mathbf{g}}(\mathbf{k})$ and $\mathbf{v}_{\mathbf{k}}$. The supercurrent density is related to the phase gradient through $\mathbf{J}\!=\!\rho_{\text{sf}}\,\mathbf{Q}$, where the superfluid stiffness is given by $\rho_{\text{sf}}= (q^*\,\hbar/m^\ast)\,n_s$. Here, $n_s$ is the Cooper-pair density, $q^{*}$ is the pair charge, and $m^{*}$ is the effective pair mass. Combining these relations yields a current-induced intrinsic magnetization as $M_z=\alpha_J J$, where $\alpha_J=\kappa/\rho_{\text{sf}}$. For an effective
anisotropic out-of-plane SOC interaction $g_z(\theta)=\lambda_0+\lambda_1\cos\theta$, we obtain
\begin{equation}
    \alpha_J = \dfrac{\mu_B}{e v_F} \frac{\lambda_1 (\lambda_R k_F)^2}
{\big[(\lambda_R k_F)^2 + \lambda_0^2\big]^{3/2}}\,~.
\label{eq:alphaJ2}
\end{equation}
This relation clearly shows that both an anisotropic out-of-plane coupling ($\lambda_1\!\neq\!0$) and Rashba interaction ($\lambda_R \neq 0$) are essential for a nonzero magnetoelectric coefficient. We anticipate the effective out-of-plane coupling to be on the order of the Zeeman energy scale. Setting $\lambda_0 = 0$, $\lambda_1 \approx 0.12$–$1.16~\mu\text{eV}$, and using $\mu_B/(e v_F)\approx 10~{\rm G\,cm^2/A}$, $E_F\approx 10-20$~meV~\cite{lubashevsky_2012,rinott_2017}, $\lambda_R k_F/E_F\approx 0.1$, Eq.~\eqref{eq:alphaJ2} yields $\alpha_J \approx 0.6-11.6~{\rm G\,cm^2/kA}$, in good agreement with the experimental results shown in Fig.~\ref{fig_coupling}(d).

In summary, we demonstrate that the superconducting condensate in mesoscopic Fe(Te,Se) rings develops an intrinsic magnetization. The effective field is reflected through the phase modulation of Little--Parks oscillations and is coupled to DC supercurrent through spin–orbit interaction. We uncover a dual flux-quantization mechanism that allows electrical tuning of fluxoid states at fixed external magnetic field, opening new routes for scalable quantum hardware and superconducting spintronics. These findings provide strong evidence for spin-polarized superconductivity at the device scale, marking a decisive step toward the detection and manipulation of exotic states in topological superconductors. 

\bibliography{REF}

\setcounter{equation}{0}
\setcounter{figure}{0}
\setcounter{table}{0}
\renewcommand{\theequation}{A\arabic{equation}}
\renewcommand{\thefigure}{A\arabic{figure}}

\begin{center}
\textbf{\large End Matter}
\end{center}

\textit{Appendix A: LP oscillations.}
The free energy of a superconducting ring is given by 
$F_n(\Phi)\propto (n-\Phi/\Phi_0)^2$, where $n$ is the winding number~\cite{littlepark_1962,littlepark_1964}. The ground state is determined by the value of $n$ that minimizes $F_n$, yielding a periodic envelope 
$F(\Phi)=\min_{n} \left[(n-\Phi/\Phi_0)^2\right]$. This piecewise-parabolic function can be expanded as a Fourier series in harmonics of the reduced flux 
$\varphi=\Phi/\Phi_0$
\begin{equation}
F(\Phi)=\frac{1}{12}+\frac{1}{\pi^2}\sum_{m=1}^{\infty}\frac{(-1)^m}{m^2}
\cos(2\pi m\,\varphi).
\end{equation}
To leading order, this results in Little–Parks oscillations of the form
$F(\Phi)\simeq F_0 - A\cos(2\pi\Phi/\Phi_0)$.\\

\textit{Appendix B: Power scaling.} The effective field is predicted to scale with the inverse of the ring size~\cite{vakaryuk_2009}. The inverse scaling of the coupling factor could originate from the stability requirement of a circulating spin supercurrent. Since the free energy associated with spin current grows logarithmically with system size~\cite{sigrist_1991}, these states are thermodynamically favored only when the ratio of spin to charge superfluid densities satisfies $\rho_{sp}/\rho_s < (1+\beta)^{-1}$, where $\beta = \text{Lw}/2\lambda^{2}$ and $\lambda$ is the London penetration depth~\cite{chung_2007}. Reducing Lw suppresses this energy cost and promotes the stability of circulating spin currents, consistent with the enhanced coupling strength in the smaller rings.\\

\textit{Appendix C: Zeeman coupling.} In the presence of Zeeman coupling, the spin texture is canted out of the plane with respect to the helicity index of each Fermi surface. Such a spin-z component scales directly with the external magnetic field and vanishes when the field is removed. Zeeman coupling would therefore introduce an explicit field dependence of the intrinsic magnetization. In contrast, our results points to an intrinsic magnetization that is independent of the external magnetic field and persists even at $B_z\!=\!0$ (see Fig.~\ref{fig_heff}(a)). Moreover, as shown in the \textcolor{cyan}{SM [Eq.~S24]}, an isotropic $s_z(\mathbf{k})$ integrates to zero magnetization, implying that Zeeman coupling cannot generate a finite out-of-plane magnetization. These observations collectively are not consistent with Zeeman-coupling origin of the effective field. From a microscopic perspective, the inclusion of an effective out-of-plane spin–orbit component $g_z(\mathbf{k})$ is justified in iron-based superconductors. Spin-resolved ARPES measurements in representative iron-based supercondcutors such as FeSe and LiFeAs provide direct evidence for out-of-plane spin polarization and explicitly link these spin components to SOC-induced effects~\cite{day_2018}. SOC in these materials is sizable and substantially reshapes the low-energy electronic structure~\cite{borisenko_2016}. The coexistence of in-plane and out-in-plane SOC has been extensively studied in non centrosymmetric and Ising superconductor contexts and is known to strongly enhance the magnetoelectric responses~\cite{lu_2015,xi_2016,Zhou_2016}.\\

\pagebreak
\widetext
\begin{center}
\textbf{\large Supplemental Materials}
\end{center}
\setcounter{equation}{0}
\setcounter{figure}{0}
\setcounter{table}{0}
\makeatletter
\renewcommand{\theequation}{S\arabic{equation}}
\renewcommand{\thefigure}{S\arabic{figure}}
\section{Theory}
\subsection{A. Mixed Pairing in a SOC System} 
In the following, we show that the pairing potential in a superconductor with spin-orbit coupling (SOC) naturally mixes singlet and triplet components. The Hamiltonian of 2D system with in-plane [\textcolor{blue}{1-3}] and out-of-plane [\textcolor{blue}{4-7}] SOC reads as
\begin{equation}
H(\mathbf k) = \xi_{\mathbf k}\sigma_0 + \lambda_R(k_x\sigma_y-k_y\sigma_x) + g_z(\mathbf k)\sigma_z~,
\label{Hamiltonian}
\end{equation} 
where $\xi_{\mathbf{k}} = (\hbar^2 k^2)/(2m)$, $\lambda_R$ is the Rashba coupling strength, and $g_z(\mathbf{k})$ encodes the out-of-plane spin--orbit interaction. We introduce the SOC vector
\begin{equation}
\mathbf g(\mathbf k) = \big(\lambda_R k_y,\,-\lambda_R k_x,\,g_z(\mathbf k)\big)~,
\end{equation}
where $g_k \equiv |\mathbf g(\mathbf k)|$, and $ \hat{\mathbf g}(\mathbf k) = \mathbf g(\mathbf k)/g_k$.
The normal-state eigenvalues are given by
\begin{equation}
E_{\mathbf k s} = \xi_{\mathbf k} + s g_k~.
\label{energies}
\end{equation} 
Here, $s=\pm$ is the helicity index. We parametrize $\hat{\mathbf g}(\mathbf k)$ using spherical angles
$\chi_{\mathbf k}$ and $\varphi_{\mathbf k}$,
\begin{equation}
\hat{\mathbf g}(\mathbf k) = (\sin\chi_{\mathbf k}\cos\varphi_{\mathbf k}, \sin\chi_{\mathbf k}\sin\varphi_{\mathbf k},\cos\chi_{\mathbf k})~,
\end{equation}
with
$\cos\chi_{\mathbf k}=g_z(\mathbf k)/g_k$ and
$\sin\chi_{\mathbf k}=\lambda_R k/g_k$.
A convenient choice of normalized helicity eigenstates are 
\begin{align}
|\mathbf k,+\rangle & = 
\begin{pmatrix}
\cos\frac{\chi_{\mathbf k}}{2}\\[2pt]
-\,i e^{i\varphi_{\mathbf k}}\sin\frac{\chi_{\mathbf k}}{2}
\end{pmatrix}~,\\
|\mathbf k,-\rangle & =
\begin{pmatrix}
-\,i e^{-i\varphi_{\mathbf k}}\sin\frac{\chi_{\mathbf k}}{2}\\[2pt]
\cos\frac{\chi_{\mathbf k}}{2}
\end{pmatrix}~.
\end{align}
For $g_z(\mathbf k)=0$ , $\chi_{\mathbf k}=\pi/2$ and the above basis reduces to the
pure-Rashba spinors. We define helicity operators as
\begin{equation}
\gamma_{\mathbf k s} = u_s(\mathbf k)\,c_{\mathbf k\uparrow} + v_s(\mathbf k)\,c_{\mathbf k\downarrow}~,
\end{equation}
with the following coefficients
\begin{align}
u_+ = \cos\frac{\chi_{\mathbf k}}{2}~,
\qquad
&v_+ = -\,i e^{i\varphi_{\mathbf k}}
       \sin\frac{\chi_{\mathbf k}}{2}~, \notag \\[3pt]
u_- = -\,i e^{-i\varphi_{\mathbf k}}
       \sin\frac{\chi_{\mathbf k}}{2}~,
\qquad
&v_- = \cos\frac{\chi_{\mathbf k}}{2}~.
\end{align}
Assuming intraband pairing in the strong SOC limit, the pairing Hamiltonian is then given by
\begin{equation}
H_\Delta
= \frac{1}{2}\sum_{\mathbf k}
\Big[
\Delta_+(\mathbf k)\,\gamma_{\mathbf k+}^\dagger\gamma_{-\mathbf k+}^\dagger
+
\Delta_-(\mathbf k)\,\gamma_{\mathbf k-}^\dagger\gamma_{-\mathbf k-}^\dagger
+ \text{h.c.}
\Big]~.
\label{eq:HD_Rashba_Ising}
\end{equation}
For the spin basis, we expand the helicity operators as
\begin{align}
\gamma_{\mathbf k s}^\dagger\gamma_{-\mathbf k s}^\dagger &= u_s^2\, c_{\mathbf k\uparrow}^\dagger c_{-\mathbf k\uparrow}^\dagger + v_s^2\, c_{\mathbf k\downarrow}^\dagger c_{-\mathbf k\downarrow}^\dagger \notag \\
& + u_s v_s \big( c_{\mathbf k\uparrow}^\dagger c_{-\mathbf k\downarrow}^\dagger
+ c_{\mathbf k\downarrow}^\dagger c_{-\mathbf k\uparrow}^\dagger\big)~.
\end{align}
Inserting the explicit expressions for $u_s$ and $v_s$ and summing over $s=\pm$ with weights $\Delta_s(\mathbf k)$, the pairing Hamiltonian can be rewritten in the spin basis as
\begin{align}
H_\Delta = \frac12 &\sum_{\mathbf{k}} \bigg\{
\psi(\mathbf{k}) 
\big(c_{\mathbf{k}\uparrow}^\dagger c_{-\mathbf{k}\downarrow}^\dagger 
     - c_{\mathbf{k}\downarrow}^\dagger c_{-\mathbf{k}\uparrow}^\dagger\big) \notag \\
     & +\mathbf{d}(\mathbf{k})\cdot \big( c_{\mathbf{k}\alpha}^\dagger (\boldsymbol{\sigma} \, i\sigma_y)_{\alpha\beta} c_{-\mathbf{k}\beta}^\dagger \big) + \text{h.c.} \bigg\}~,
\end{align}
where 
\begin{align}
\psi(\mathbf k) &= \tfrac12\big[\Delta_+(\mathbf k)+\Delta_-(\mathbf k)\big]~,
\\[4pt]
\mathbf d(\mathbf k)
&= \tfrac12\big[\Delta_+(\mathbf k)-\Delta_-(\mathbf k)\big]\,
\hat{\mathbf g}(\mathbf k)~.
\end{align}
Accordingly, the gap function in the spin basis takes the following compact form
\begin{equation}
\hat\Delta(\mathbf k)
=
\big[\psi(\mathbf k)
+ \mathbf d(\mathbf k)\cdot\boldsymbol{\sigma}\big]\,
i\sigma_y~.
\label{eq:Delta_Rashba_Ising}
\end{equation}
The first term $\psi(\mathbf{k}) i\sigma_y$ is \emph{antisymmetric} under exchange of the two spin indices $(i\sigma_y)^T = -\, i\sigma_y$, and therefore corresponds to the conventional spin-singlet state with total spin of $S=0$. In contrast, $\sigma_i\, i\sigma_y$ is \emph{symmetric} under spin exchange $(\sigma_i i\sigma_y)^T = +\, \sigma_i i\sigma_y$, and hence the second term $\mathbf{d}(\mathbf{k})\cdot\boldsymbol{\sigma}\, i\sigma_y$ is a spin-triplet component with total spin of $S=1$. Although spin--orbit interaction mixes spin and momentum in a way that the total spin is no longer a good quantum number, the symmetry of the gap function in the spin space remains well-defined. Consequently, the pairing potential can still be classified into singlet and triplet sectors according to the antisymmetric or symmetric structure of the gap function~[\textcolor{blue}{8-9}].

\subsection{B. Magnetoelectric Coefficient}
In this section, we consider a minimal model with SOC in which a uniform supercurrent drives an out-of-plane spin polarization of the condensate. This response can be equivalently described as an intrinsic magnetic field proportional to the current density, $B_{\mathrm{eff},z}=\alpha_J J$. Within this framework, we derive a microscopic expression for the coupling coefficient $\alpha_J$. Using the model Hamiltonian for a SOC superconductor [Eq.~\ref{Hamiltonian}], the expectation value of the spin operator in helicity band $s$ is determined by
\begin{equation}
\langle \boldsymbol{\sigma} \rangle_{s}(\mathbf{k})
= s\, \hat{\mathbf{g}}(\mathbf{k})
= s\, \frac{\mathbf{g}(\mathbf{k})}{g(\mathbf{k})}~,
\end{equation}
with the out-of-plane component given by
\begin{equation}
s_z^{s}(\mathbf{k}) = \langle \sigma_z \rangle_s(\mathbf{k}) = s\,\frac{g_z(\mathbf{k})}{g(\mathbf{k})}~.
\label{eq:sz-def}
\end{equation}

When a uniform supercurrent flows through the system, the condensate develops a finite center-of-mass momentum $\mathbf{Q}$, which can be expressed as a uniform phase gradient of the order parameter $\psi(\mathbf{r}) = |\psi| e^{i(\phi_0 + \mathbf{Q}\cdot \mathbf{r})}$. The supercurrent density is directly proportional to the phase gradient through
\begin{equation}
\mathbf{J} = \rho_{\text{sf}}\, \mathbf{Q}~,
\label{eq:London}
\end{equation}
where $\rho_{\text{sf}}=(q^* \hbar/m^*)\,n_s$ is the superfluid stiffness, $n_s$ is the Cooper-pair density, $q^{*}=2e$ is the pair charge, and $m^{*}$ is the effective pair mass. The finite center-of-mass momentum shifts the quasiparticle energies $E_{k} \rightarrow E_{k} + \delta E_{k}$, where the Doppler correction is linear in $\mathbf{Q}$~[\textcolor{blue}{10}]. In the linear-response regime, this leads to the following expression for the Doppler shift 
\begin{equation}
\delta E_{\mathbf{k},s} = \hbar\, \mathbf{v}_{\mathbf{k},s} \!\cdot\! \mathbf{Q}~,
\label{eq:Doppler}
\end{equation}
where the group velocity is given by
\begin{equation}
\mathbf{v}_{\mathbf{k},s} = \hbar^{-1} \nabla_{\mathbf{k}} E_{\mathbf{k},s}~.
\end{equation}
This energy shift, in turn, perturbs the equilibrium distribution function, producing a change in the occupation of states close to the Fermi surface
\begin{equation}
\delta n_{\mathbf{k},s} = - \left( \frac{\partial f}{\partial E} \right)_{E = E_{\mathbf{k},s}} \delta E_{\mathbf{k},s}~,
\label{eq:delta-n}
\end{equation}
where $f(E)$ is the Fermi distribution function.

For each Fermi surface, the current-induced out-of-plane magnetization density can be evaluated through~[\textcolor{blue}{11}]
\begin{equation}
M_z(\mathbf{Q}) = \mu_B \sum_{\mathbf{k}} s_z(\mathbf{k})\, \delta n_{\mathbf{k}}~,
\label{eq:Mz-def}
\end{equation}
where $s_z(\mathbf{k})$ is given by Eq.~\eqref{eq:sz-def}. Inserting Eq.~\eqref{eq:delta-n} results in 
\begin{equation}
M_z(\mathbf{Q}) = \hbar \mu_B \sum_{\mathbf{k}} s_z(\mathbf{k})\,\big(\mathbf{v}_{\mathbf{k}} \!\cdot\! \mathbf{Q}\big)
\left( - \frac{\partial f}{\partial E} \right)_{E = E_{\mathbf{k}}}~.
\label{eq:Mz-k-sum}
\end{equation}
We project the velocity along the current direction, and for concreteness we take
$\mathbf{Q} = Q \hat{x}$ so that
\begin{equation}
\mathbf{v}_{\mathbf{k}} \!\cdot\! \mathbf{Q}
\simeq v_{F}\, Q \cos\theta~,
\end{equation}
where we used $v_k \approx v_{F}$. Equation~\eqref{eq:Mz-k-sum} then yields
\begin{equation}
M_z(Q) = \hbar\mu_B  Q N_F v_{F} \oint_{\mathrm{FS}} \frac{d\theta}{2\pi}\, s_z(\theta)\, \cos\theta~.
\label{eq:Mz-Q-general}
\end{equation}
where $N_F$ is the density of state near Fermi surface. We introduce a dimensionless \textit{magnetoelectric form factor}
\begin{equation}
\eta \equiv \oint_{\mathrm{FS}} \frac{d\theta}{2\pi}\,s_z(\theta)\, \cos\theta~,
\label{eq:eta-def}
\end{equation}
The out-of-plane magnetization density \eqref{eq:Mz-Q-general} then reads
\begin{equation}
M_z(Q) = \kappa\, Q~.
\label{eq:Mz-kappa}
\end{equation}
Here, $\kappa$ is the \textit{magnetoelectric kernel} given by
\begin{equation}
\kappa = \hbar \mu_B N_F\, v_{F}\, \eta~.
\label{eq:Mz-kappa2}
\end{equation}
The current-induced out-of-plane magnetization is obtained by combining Eqs~\eqref{eq:London} and \eqref{eq:Mz-kappa}
\begin{equation}
M_z = \frac{\kappa}{\rho_{\text{sf}}}\, J~,
\label{eq:Mz-from-I}
\end{equation}
and the intrinsic magnetoelectric coupling factor reads as
\begin{equation}
\alpha_J =  \frac{\kappa}{\rho_{\text{sf}}} = \frac{\mu_B \hbar}{\rho_{\text{sf}}} N_F\, v_{F}\, \eta~.
\label{eq:alpha-final}
\end{equation}
The sign and magnitude of $\alpha_J$ are controlled by the spin–orbit texture and the relative superfluid weight on the two helicity-resolved Fermi surfaces. For an isotropic out-of-plane SOC, symmetry enforces a cancellation of the Fermi-surface contributions, yielding $\eta=0$. Any departure from isotropy, even infinitesimal, lifts this cancellation and renders $\eta$ finite. We adopt an effective anisotropic out-of-plane SOC interaction as
\begin{equation}
g_z(\mathbf{k}) \equiv g_z(\theta) = \lambda_0 + \lambda_1 \cos\theta~.
\label{eq:gz-ansatz}
\end{equation}
Here, $\theta$ parametrizes the direction on the Fermi surface, $\mathbf{k} = k_F(\cos\theta,\sin\theta)$. For a fixed $k_F$, the in-plane Rashba field has constant magnitude $\lambda_R k_F$. Using $g(\theta) = \sqrt{\lambda_R^2 k_F^2 + g_z(\theta)^2}$, the spin-$z$ expectation on helicity band $s=\pm$ is given by
\begin{equation}
s_z(\theta) = s\,\frac{g_z(\theta)}{g(\theta)} = s\,\frac{\lambda_0 + \lambda_1 \cos\theta} {\sqrt{\lambda_R^2 k_F^2 + \big(\lambda_0 + \lambda_1 \cos\theta\big)^2}}~.
\label{eq:sz-theta}
\end{equation}
Substituting this equation into Eq.~\eqref{eq:eta-def}, the magnetoelectric form factor reads as
\begin{equation}
\eta
\simeq
s\,\frac{\lambda_1 (\lambda_R k_F)^2}
{2\big[(\lambda_R k_F)^2 + \lambda_0^2\big]^{3/2}}.
\label{eq:eta-final}
\end{equation}

Equation~\eqref{eq:eta-final} makes several key points transparent. First, for an isotropic out-of-plane SOC ($\lambda_1=0$), the magnetoelectric form factor vanishes. Second, in the absence of Rashba interaction ($\lambda_R=0$), $\eta$ also vanishes, which is consistent with the fact that spin-momentum locking is an essential key for the current-induced magnetoelectric response. Finally, the two helicity-resolved bands contribute with opposite signs, such that $\mathrm{sign}(\eta_+)=-\mathrm{sign}(\eta_-)$. This sign structure underlies the observed reversal of the intrinsic field at flip current $I_F$ discussed in the paper. Inserting Eq.~\eqref{eq:eta-final} into Eq.~\eqref{eq:alpha-final}, we obtain a microscopic expression for the coupling factor
\begin{equation}
\label{eq:final2}
\alpha_J = \dfrac{\mu_B}{e v_F} \frac{\lambda_1 (\lambda_R k_F)^2} {\big[(\lambda_R k_F)^2 + \lambda_0^2\big]^{3/2}}~.
\end{equation}
We now estimate the current-induced magnetoelectric coefficient $\alpha_J$ for a Rashba systems with an effective anisotropic out-of-plane SOC. Intuitively, we anticipate the effective out-of-plane coupling to be on the order of the Zeeman energy scale. Assuming $\lambda_0 = 0$ and taking $\lambda_1 \simeq 0.12$–$1.16~\mu\text{eV}$, corresponding to the Zeeman energy at a field range of 10–100 G, and using $\mu_B/(e v_F)\simeq 10~{\rm G\,cm^2/A}$, $E_F\simeq 10-20$~meV~[\textcolor{blue}{12,13}], and $\lambda_R k_F/E_F\simeq 0.1$, Eq.~\eqref{eq:final2} yields $\alpha_J \approx 0.6\text{--}11.6~{\rm G\,cm^2/kA}$, in good agreement with the experimental results shown in Fig.~4(d) in the main manuscript.\\

\clearpage
\newpage

\begin{table*}[t]
    \caption{\textbf{Characteristic scales of mesoscopic Fe(Te,Se) ring devices}. All devices share a similar layout with 12 identical rings [Fig.~1(b)]. $t_\text{FTS}$ and $t_\text{hBN}$ denote the thicknesses of the Fe(Te,Se) and hBN layers, respectively. $L_i$ and w are the inner-edge length and wall width of the rings, respectively, and $L$w denotes the effective ring size, corresponding to the two-dimensional area over which the superfluid circulates ($L\!=\!L_i\!+\!\text{w}$). The effective flux area is calculated as $A_\text{eff}\!=\! L^2\left[1 + (\text{w}/L)^2\right]$, where the second term accounts for the finite wall width. $\Delta B_g$ is the expected flux periodicity estimated from the ring geometry, $\Delta B_g = \Phi_0/A_\text{eff}$, and $\Delta B_m$ is the measured field periodicity extracted from the FFT of the MR oscillations. The coupling factor $\alpha$ is obtained from the constant-phase lines as discussed in Fig.~\ref{figsi_alldev}, and $\alpha_J$ is calculated through $\alpha_J = \alpha\,(\text{w}\,t_{_\text{FTS}})$.}
    \label{tabsi_1}
    \begin{tabular}{ccccccccccc}
        \hline
          Device & $t_\text{\tiny{FTS}}$~(nm) & $t_\text{\tiny{hBN}}$~(nm) & $L_{i}~(nm)$ & $\text{w}$~(nm) & $L\text{w}$~($\mu m^2$) & $A_\text{eff}$~($\mu m^2$) & $\Delta B_\text{g}$~(G) & $\Delta B_\text{m}$~(G) & $\alpha$~(G/$\mu$A) & $\alpha_J$~(G cm$^2$/$\mu$A) \\
        \hline 
        1 & 63 & 22 & 406$\pm$19 & 118$\pm$17 & .06$\pm$.01 & .29$\pm$.03 & 72$\pm$7 & 65.8 & 17.2$\pm$4.4 & 1.3$\pm$0.3\\
        2 & 63 & 22 & 415$\pm$15 & 174$\pm$21 & .10$\pm$.02 & .38$\pm$.04 & 55$\pm$5 & 54.1 & 11.6$\pm$3.9 & 1.3$\pm$0.4\\
        3 & 86 & 30 &  454$\pm$16 & 227$\pm$15 & .15$\pm$.01 & .52$\pm$.03 & 40$\pm$3 & 47.6 & 13.1$\pm$1.8 & 2.6$\pm$0.3\\
        4 & 105 & 65 &  401$\pm$12 & 295$\pm$16 & .21$\pm$.02 & .57$\pm$.04 & 36$\pm$2 & 49.3 & 8.9$\pm$0.5 & 2.8$\pm$0.2\\
        5 & 33 & 27 &  195$\pm$14 & 97$\pm$12 & .03$\pm$.00 & .09$\pm$.01 & 218$\pm$3 & 232.6 & 75.3$\pm$10.6 & 2.4$\pm$0.3\\
        6 & 45 & 55 &  365$\pm$2 & 127$\pm$3 & .06$\pm$.01 & .25$\pm$.00 & 80$\pm$1 & 83.0 & 21.5$\pm$3.5 & 1.2$\pm$0.2\\
        \hline
    \end{tabular} 
\end{table*}

\begin{figure*}
    \centering
    \includegraphics[width=0.9\textwidth]{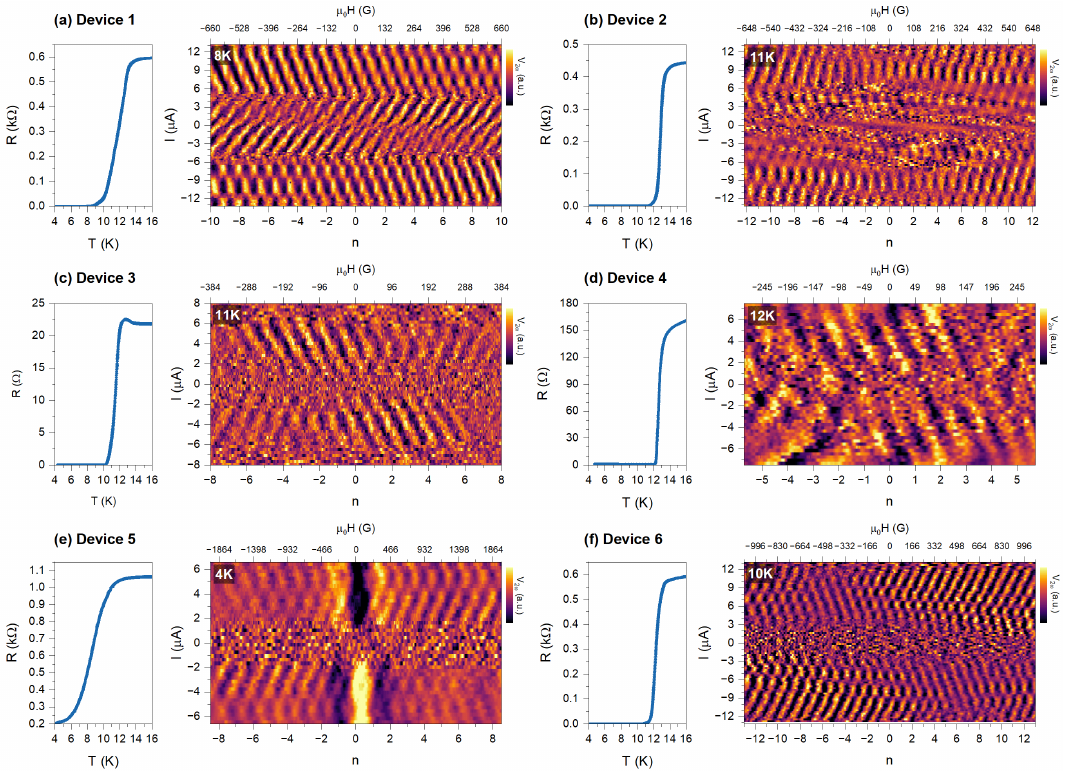}
    \caption{\textbf{Reproducibility of the Little--Parks phase.} Superconducting transition and normalized MR oscillations as functions of the winding number (external magnetic field) and the applied DC current for six representative devices with different ring sizes. The characteristic length scales, field periodicities, and magnetoelectric coupling factors of these devices are summarized in Table~\ref{tabsi_1}. For each device, the winding number is defined as $n = \mu_0 H / \Delta B_m$, where $\Delta B_m$ is the field periodicity extracted from the FFT of the corresponding LP oscillations. MR data are collected at temperatures corresponding to the onset of the resistive state, except for device 5, where measurements are performed at 4 K (base temperature) with a residual resistance ratio of $R(4~\mathrm{K})/R(16~\mathrm{K}) \approx 0.2$. All devices with effective ring sizes $Lw < 0.2~\mu\mathrm{m}^2$ exhibit similar transport characteristics, including a global current–field symmetry, the emergence of HIW states arising from the splitting of IW modes, and a current-driven LP phase $\theta_{LP}(I)$. The absence of these features in larger rings can be attributed to the instability of spin-polarized circulating supercurrents in extended geometries (see Appendix B). The coupling factor associated with the intrinsic effective field are extracted from the slope of constant-phase lines and are summarized in Table~\ref{tabsi_1}.}
    \label{figsi_alldev}
\end{figure*}

\clearpage
\newpage

\section{References}
\begingroup 
  \renewcommand{\labelenumi}{[\theenumi]} 
  \begin{enumerate}
\item Y. A. Bychkov and ´E. I. Rashba, Properties of a 2d elec-
tron gas with lifted spectral degeneracy, JETP lett 39,
78 (1984). 
\item T. Oguchi and T. Shishidou, The surface rashba effect:
a k.p perturbation approach, Journal of Physics: Con-
densed Matter 21, 092001 (2009). 
\item P. Zhang, K. Yaji, T. Hashimoto, Y. Ota, T. Kondo,
K. Okazaki, Z. Wang, J. Wen, G. D. Gu, H. Ding, et al.,
Observation of topological superconductivity on the sur-
face of an iron-based superconductor, Science 360, 182
(2018).
\item R. P. Day, G. Levy, M. Michiardi, B. Zwartsenberg,
M. Zonno, F. Ji, E. Razzoli, F. Boschini, S. Chi, R. Liang,
P. K. Das, I. Vobornik, J. Fujii, W. N. Hardy, D. A. Bonn,
I. S. Elfimov, and A. Damascelli, Influence of spin-orbit
coupling in iron-based superconductors, Phys. Rev. Lett.
121, 076401 (2018).
\item B. T. Zhou, N. F. Yuan, H.-L. Jiang, and K. T. Law, Ising
superconductivity and majorana fermions in transition-
metal dichalcogenides, Physical Review B 93, 180501
(2016).
\item J. Lu, O. Zheliuk, I. Leermakers, N. F. Yuan, U. Zeitler,
K. T. Law, and J. Ye, Evidence for two-dimensional
ising superconductivity in gated MoS2, Science 350, 1353
(2015).
\item X. Xi, Z. Wang, W. Zhao, J.-H. Park, K. T. Law,
H. Berger, L. Forr´o, J. Shan, and K. F. Mak, Ising pairing
in superconducting NbSe2 atomic layers, Nature Physics
12, 139 (2016).
\item P. Frigeri, D. Agterberg, A. Koga, and M. Sigrist, Super-
conductivity without inversion symmetry: MnSi versus
CePt3Si, Physical review letters 92, 097001 (2004).
\item V. Cvetkovic and O. Vafek, Space group symmetry, spin-
orbit coupling, and the low-energy effective hamiltonian
for iron-based superconductors, Phys. Rev. B 88, 134510
(2013).
\item V. M. Edelstein, Magnetoelectric effect in polar super-
conductors, Physical review letters 75, 2004 (1995).
\item S. Fujimoto, Electron correlation and pairing states in
superconductors without inversion symmetry, Journal of
the Physical Society of Japan 76, 051008 (2007).
\item Y. Lubashevsky, E. Lahoud, K. Chashka, D. Podolsky,
and A. Kanigel, Shallow pockets and very strong coupling
superconductivity in FeSexTe1-x, Nature Physics 8, 309
(2012).
\item S. Rinott, K. Chashka, A. Ribak, E. D. Rienks, A. Taleb-
Ibrahimi, P. Le Fevre, F. Bertran, M. Randeria, and
A. Kanigel, Tuning across the BCS-BEC crossover
in the multiband superconductor Fe1+y SexTe1-x: An
angle-resolved photoemission study, Science advances 3,
e1602372 (2017).
 \end{enumerate}
\endgroup
\end{document}